# Behaviour of One-Site Entanglement and Reduced Fidelity in an Anisotropic Spin-2 chain

**Victoria Sharmila Gomes**[1,2], **Amit Tribedi**[2], **Subhrajyoti Dey**[1]
[1]*Department of Physics, School of Basic Science, Swami Vivekananda University, Barrackpore, West Bengal-700121, India*
[2]*Department of Physics, Sushil Kar College, Ghoshpur, Champahati, South 24 Parganas, West Bengal, India.*
Email: *amittribedi@outlook.com*, *victoriagomes0412@gmail.com*

**Abstract**

Utilizing the matrix product formalism, we have studied the variation of entanglement and fidelity measures in the MP ground states of a generic anisotropic spin-2 chain with nearest neighbour interactions and shared symmetries. These MP states represent the exact Ground State solutions of the system and display distinctive characteristics determined by a maximum of three parameters exhibiting one antiferromagnetic Haldane phase (AFMH), together with a secondary weak antiferromagnetic (WAFM) and a weak ferromagnetic phase (WFM). Using Transfer Matrix Method (TM), we have obtained nontrivial features in the variation of Quantum Correlation measures and QIT Measures around the VBS point (for an 'acritical' choice of the parameters) and a Quantum Critical Point (for a 'critical' choice of the parameters) of the AFMH spin liquid phase of the system.

Keywords: Spin-2 chains, Haldane phases, Quantum Fluctuations, Matrix Product Approach, Entanglement, Fidelity

**Introduction**

Haldane's conjecture [1,2] sparked interest in low-dimensional systems with different spin values, leading to the study of materials representing spin chains with spin values greater than 1/2. Due to quantum fluctuations, mean-field theories are unreliable for low-dimensional models, making exact results crucial for exploring phases and properties. There are, however, some alternative methods that provide a systematic way to construct exact ground states (GS) for various quantum systems beyond one-dimensional spin chains. One such powerful method is the use of 'optimal' GSs (OGS).

Since only a few models can be exactly solved, such as through Bethe-Ansatz [3], alternative methods like numerical techniques (e.g. DMRG) [4,5] and powerful analytical approaches are necessary. This study utilizes the concept of OGS [6–8] of systematically constructing exact GSs, applicable not only to one-dimensional spin systems but also to quantum spin systems in any dimension [9,10], Hubbard models [8], or stochastic processes [11]. The matrix product (MP) formalism has also been employed in spin-1 quantum chains, showcasing its flexibility in discovering precise GSs and analyzing their correlation functions [12]. Additionally, the accurate GSs of spin-1/2 and spin-3/2 quantum chains have been proven to disrupt rotational and translational symmetries, with explicit correlation functions offered for these models [13]. Within the framework of the extended Hubbard model, precise GSs display spin correlations located at long ranges, and in cases with SU (2) symmetry, these states coincide with higher spin states, including completely ferromagnetic states [14]. The research by Fannes et al. [15] on translation-invariant states for infinite quantum spin chains, particularly for integer spins, further bolsters the existence of exponential decay in correlation functions and offers detailed computations for these characteristics. Batchelor and Yung's investigation of valence-bond (VB) GSs in quantum spin chains, utilizing q-deformation of SU (2), expands these discoveries to anisotropic models [16]. Tu and Sanz's real-space renormalization-group technique provides an extensive collection of precise solutions for SU (2) symmetric quantum spin chains with spins up to 5, identifying matrix-product states as well as partially ferromagnetic states as GSs [17]. Finally, Takano *et al.* explore the combined diamond chain, where various GSs arise from different exchange parameters, including the Haldane state and spin cluster states [18]. As a whole, these studies emphasize the diverse and complex nature of precise GSs in spin-2 and related quantum spin chains.

Recently, there has been a significant amount of research on quantum phase transitions (QPTs) in spin systems, focusing on established measures from quantum information theory (QIT). Several entanglement measures have been discovered that exhibit distinct characteristics at the quantum critical point [19–25]. Research has demonstrated that a first-order quantum phase transition (QPT) [22] is typically indicated by a discontinuity in the first derivative of the ground-state energy. This discontinuity is also exhibited by measures of entanglement across a bipartition of the system. Also, a continuous phase transition is identified with a discontinuity or divergence in the first or higher derivatives of the entanglement measure, which corresponds to a discontinuity or divergence in the second or higher derivatives of the GS energy.

Fidelity (*F*), an idea borrowed from QIT, is depicted as the overlay between GSs of two parameter values separated by a differentially small amount. What sets fidelity apart is its ability to identify QPTs without needing prior knowledge about the system's symmetries or order parameters. A sharp drop in fidelity at a critical point indicates a significant change in the GS wave function. Researchers have linked the reduction in fidelity with the vanishing of the 'proverbial' excitation gap, particularly in MP states, and have used fidelity derivatives for effective extraction of critical exponents. By measuring fidelity, researchers have successfully predicted QPTs in the Bose-Hubbard model, surpassing traditional methods like entanglement assessments. In particular situations, like the antiferromagnetic Heisenberg spin chain with various interactions, the authenticity of the first excited state, rather than the GS, demonstrates to be a crucial indicator of QPTs. Newly established concepts such as reduced fidelity (RF) [26–32] and reduced fidelity susceptibility (RFS) have recently been developed. This is particularly relevant in cases where the global GS is not accessible for a physical system. The term "RF" pertains to a part of the whole system and is characterized as the intersection between the density matrices of the said part (obtained by taking partial trace over the rest of the system) corresponding to two different Hamiltonian parameters having a differential difference and RFS acts as a 'response function' associated with the double derivative of the RF. In spin models such as the Lipkin-Meshkov-Glick model [29,30], the transverse field Ising model in 1d [33], and the spin-1/2 dimerized Heisenberg chain [31], QPTs have been investigated using the RF and RFS measurements [32].

MPGS is a method used to construct GSs for spin systems, such as the spin-2 chain, by combining matrices representing single-spin states at different sites. These matrices are multiplied using a tensor product operation to generate the overall state of the system. This approach allows for the creation of structured and computationally tractable GSs for complex quantum systems. The use of matrix products simplifies the representation of complex GSs and enables the calculation of GS properties for various operators in the system. We can describe how matrices are used to represent single-spin states at a specific site *i* in a system. When two such matrices are multiplied together, the resulting matrix product is defined using the tensor product symbol $\otimes$ to combine the spin states from each matrix, as

$$(g^{(i)} \cdot g^{(i+1)})_{\mu\nu} = \sum_k g^{(i)}_{\mu k} \otimes g^{(i+1)}_{k\nu} \qquad (1)$$

This MP representation is essential for constructing MPGS regarding the spin-2 chain system. In respect to spin chains, an MPGS is a fundamental state that satisfies certain conditions for each element in the chain. The MPGS is represented as a product of matrices, and to be an optimal GS, it must fulfil the specific equations involving the interaction between neighbouring elements in the chain. An MPGS is a special type of global GS (GGS) for a 1-D spin system of size *L*, particularly when considering PBCs, can be written as,

$$|\psi_0\rangle = \mathrm{tr}(g^{(1)} \cdot g^{(2)} \cdot g^{(3)} \cdots g^{(L)}), \qquad (2)$$

where tr represents the "trace over the matrix space", i.e. $\sum_\mu (g(1) \cdot g(2) \cdot g(3) \cdots g(L))_{\mu\mu}$. This representation allows for a concise description of the GS of the spin chain system under consideration. In the context of MPGSs in spin chains, for an MPGS to be an optimal GS, it must satisfy the condition $h_{k,k+1}|\psi_0\rangle = 0$ for all *k*. This condition simplifies to

$$h_{k,k+1}(g_{(k)} \cdot g_{(k+1)})_{\mu\nu} = 0, \text{ for all } k, \mu, \nu. \qquad (3)$$

This indicates that each element of $(g_{(k)} \cdot g_{(k+1)})$ is a local GS of $(h_{k,k+1})$. This requirement ensures that the MPGS is a GSS.

The system being analysed in this article is a linear arrangement of spin-2 particles. Spin-2 particles can be conceptualized as quantum spins linked to physical entities, such as ions or atoms, organized in a linear manner. An illustrative case is the spin-2 1D system found in $MnCl_3$ (bipy) compound, where 'bipy' stands for $C_{10}H_8N_2$. In this magnetic system, $Mn^{3+}$ ions behave as spin-2 particles at each site, the magnetic interaction being generated by the orbital overlap of bonds among Manganese and Chlorine atoms. Experimental observations by Granroth et al. [34] indicate AFM characteristics with correlations that die fast.

The examination of precise GSs in spin-2 chains uncover a wide range of phases and characteristics, mainly investigated using the matrix product technique. Ahrens et al [35] have discovered all OGSs for general anisotropic spin-2 chains with interactions among nearest neighbours (NN), pinpointing three distinct phases- one antiferromagnetic Haldane phase. (AFMH), one weak antiferromagnetic phase (WAFM), and one weak ferromagnetic phase (WFM). These antiferromagnetic phases are distinguished as spin liquids with correlations that diminish in exponential fashion. The MPGS, in this case, can be shown to depend on up to three mutually independent parameters and can be used to calculate physical properties using a TM approach [20]. Ahrens *et al* [35] conducted an extensive study of MPGS in a one-dimensional spin-2 chain. The GS of the said chain system exhibits either nondegeneracy or finite degeneracy and adheres to PBCs. 5 MPGSs with interesting structures have been identified, showcasing distinct features. These MPGSs exhibit novel and distinctive quantum correlation (assessed by entanglement, Fidelity etc) properties, contributing to our understanding of GS configurations in quantum systems.

In this research, we examine the behaviours of some entanglement measures and the reduced Fidelity measures in the Spin-2 chain system for two specific range of different parameters, especially around the isotropic point and degeneracy point (*a=0*) of the said system in an acritical parameter range and around a quantum critical point, which belongs to another range of values of the said parameters. We will compare the results to examine the variation of the QIT measures across the said points and the sensitivity and efficiency of the same regarding signalling changes in the GS properties.

**The AFMH Phase & The Measures**

The Haldane-Antiferromagnet-A phase is of the Spin-2 chain depicted by the MPGS

$$|\Psi_0\rangle(a, x, \gamma) = \text{tr}\left(\prod_i^L g^{(i)}\right) \qquad (6)$$

where,

$$g^{(i)} = \begin{pmatrix} |0\rangle i & \sqrt{x}|1\rangle i & a|2\rangle i \\ \sqrt{x}|\bar{1}\rangle i & \gamma|0\rangle i & \sqrt{x}|1\rangle i \\ a|\bar{2}\rangle i & \sqrt{x}|\bar{1}\rangle i & |0\rangle i \end{pmatrix} \qquad (4)$$

is defined by a specific set of matrices with 3 continuous parameters *a, x,* and *γ*. The GS exhibits AFM behaviour with vanishing single-site magnetizations, leading to fluctuations within a specific range described by the square of the fluctuation $(\Delta S^z)^2 \in [0, 4]$. The uniform product $\Psi_0(a, x, \gamma)$ encompasses matrices $g^{(i)}$ that depend on continuous parameters *a, x*, and *γ*, and consists of spin-2 basis states $(|\bar{2}\rangle, |\bar{1}\rangle, |0\rangle, |1\rangle, |2\rangle)$. This model spans a 12-parameter space, which must adhere to specific inequalities. The GS is governed by the values of parameters *a, x,* and *γ,* and it exhibits a distinct and singular characteristic [22,36].

Keeping *x* and *γ* fixed at the values -3 and -2, if we vary *a*, the ground state stays separate as long as *a* is not equal to zero. The isotropic state is characterized by precise parameter values (e.g., *x* = -3, *a* = √6, *γ* = -2) [37] where the system changes it's nature of anisotropy from easy-plane to easy-axis, and the values of the correlation lengths become same and cross each other. The GS stays separate as long as *a* is not nonzero. However, when the value of *a* is equal to *0*, the matrix product Ansatz (MPA) no longer has a unique solution, leading to an increase in degeneracy as the size of the system grows. Both the correlation lengths, longitudinal and transverse, remain finite for all values of *a*, making the system acritical for that particular choice of the parameters. The GS exhibits AF characteristics, resulting in the absence of magnetizations at specific sites ($\langle \hat{S}^x \rangle \equiv \langle \hat{S}^y \rangle \equiv \langle \hat{S}^z \rangle \equiv 0$). Fluctuations are defined by the square of ($\Delta S^z$), which is limited to values between 0 and 4.

However, if we fix *x* and *γ* at the values 0 and 1 respectively, we get another type of GS which exhibits solely different properties. There is a point *a* = √2, at which the correlation lengths cross each other, but that is not the isotropic VBS point like the previous case. On top of that, both the correlation lengths diverge at *a=0,* which is undoubtedly a quantum critical point at which the gap Δ from the first excited state vanishes.

We have used the one-site von Neumann entropy (OSVNE) as the quantum correlation measure. The OSVNE, defined as,

$$S(i) = -\text{Tr}\rho(i)\log_2 \rho(i) \qquad (5)$$

Which is also known to be a good indicator of a QPT [20,22,24,25,37] It provides a measure of how a single spin at the site *i* is entangled with the rest of the system. The reduced density matrix $\rho(i)$, obtained by tracing out all the spins except the one at site *i* from the full GS density matrix of the system, can be calculated using the transfer matrix method (TMM) [38,38–41].

The general form of the matrix $f$ ($= g \otimes (g)^\dagger$) for one single site is given by

$$f = \begin{pmatrix} |0\rangle\langle 0| & \sqrt{x}|0\rangle\langle 1| & a|0\rangle\langle 2| & \sqrt{x}|1\rangle\langle 0| & x|1\rangle\langle 1| & a\sqrt{x}|1\rangle\langle 2| & a|2\rangle\langle 0| & a\sqrt{x}|2\rangle\langle 1| & a^2|2\rangle\langle 2| \\ \sqrt{x}|0\rangle\langle \bar{1}| & \gamma|0\rangle\langle 0| & \sqrt{x}|0\rangle\langle 1| & x|1\rangle\langle \bar{1}| & \gamma\sqrt{x}|1\rangle\langle 0| & x|1\rangle\langle 1| & a\sqrt{x}|2\rangle\langle \bar{1}| & \gamma|2\rangle\langle 0| & a\sqrt{x}|2\rangle\langle 1| \\ a|0\rangle\langle \bar{2}| & \sqrt{x}|0\rangle\langle \bar{1}| & |0\rangle\langle 0| & a\sqrt{x}|1\rangle\langle \bar{2}| & x|1\rangle\langle \bar{1}| & \sqrt{x}|1\rangle\langle 0| & a^2|2\rangle\langle \bar{2}| & a\sqrt{x}|2\rangle\langle \bar{1}| & a|2\rangle\langle 0| \\ \sqrt{x}|\bar{1}\rangle\langle 0| & x|\bar{1}\rangle\langle 1| & a\sqrt{x}|\bar{1}\rangle\langle 2| & \gamma|0\rangle\langle 0| & \gamma\sqrt{x}|0\rangle\langle 1| & a\gamma|0\rangle\langle 2| & \sqrt{x}|1\rangle\langle 0| & x|1\rangle\langle 1| & a\sqrt{x}|1\rangle\langle 2| \\ x|1\rangle\langle \bar{1}| & \gamma\sqrt{x}|\bar{1}\rangle\langle 0| & x|\bar{1}\rangle\langle 1| & \gamma\sqrt{x}|0\rangle\langle \bar{1}| & \gamma^2|0\rangle\langle 0| & \gamma\sqrt{x}|0\rangle\langle 1| & x|1\rangle\langle \bar{1}| & \gamma\sqrt{x}|1\rangle\langle 0| & x|1\rangle\langle 1| \\ a\sqrt{x}|\bar{1}\rangle\langle \bar{2}| & x|\bar{1}\rangle\langle \bar{1}| & \sqrt{x}|\bar{1}\rangle\langle 0| & a\gamma|0\rangle\langle \bar{2}| & \gamma\sqrt{x}|0\rangle\langle \bar{1}| & \gamma|0\rangle\langle 0| & a\sqrt{x}|1\rangle\langle \bar{2}| & x|1\rangle\langle \bar{1}| & \sqrt{x}|1\rangle\langle 0| \\ a|\bar{2}\rangle\langle 0| & a\sqrt{x}|\bar{2}\rangle\langle 1| & a^2|\bar{2}\rangle\langle 2| & \sqrt{x}|\bar{1}\rangle\langle 0| & x|\bar{1}\rangle\langle 1| & a\sqrt{x}|\bar{1}\rangle\langle 2| & |0\rangle\langle 0| & \sqrt{x}|0\rangle\langle 1| & a|0\rangle\langle 2| \\ a\sqrt{x}|\bar{2}\rangle\langle \bar{1}| & a\gamma|\bar{2}\rangle\langle 0| & a\sqrt{x}|\bar{2}\rangle\langle 1| & x|\bar{1}\rangle\langle \bar{1}| & \gamma\sqrt{x}|\bar{1}\rangle\langle 0| & x|\bar{1}\rangle\langle 1| & \sqrt{x}|0\rangle\langle \bar{1}| & \gamma|0\rangle\langle | & \sqrt{x}|0\rangle\langle 1| \\ a^2|\bar{2}\rangle\langle \bar{2}| & a\sqrt{x}|\bar{2}\rangle\langle \bar{1}| & a|\bar{2}\rangle\langle 0| & a\sqrt{x}|\bar{1}\rangle\langle \bar{2}| & x|\bar{1}\rangle\langle \bar{1}| & \sqrt{x}|\bar{1}\rangle\langle 0| & a|0\rangle\langle \bar{2}| & \sqrt{x}|0\rangle\langle \bar{1}| & |0\rangle\langle 0| \end{pmatrix} \qquad (7)$$

Thus, the transfer matrix *F* for the same site has the general form

$$F = \begin{pmatrix} 1 & 0 & 0 & 0 & x & 0 & 0| & 0 & a^2 \\ 0 & \gamma & 0 & 0 & 0 & x & 0 & 0 & 0 \\ 0 & 0 & 1 & 0 & 0 & 0 & 0| & 0 & 0 \\ 0| & 0 & 0 & \gamma & 0 & 0 & 0 & x & 0| \\ 0 & 0 & x & 0 & \gamma^2 & 0 & 0 & 0 & x \\ 0 & x & 0| & |0 & 0 & \gamma & 0 & 0 & 0 \\ 0 & 0 & 0 & 0 & 0| & 0 & 1 & 0 & 0 \\ 0 & 0 & 0 & x & 0 & 0 & 0| & \gamma & 0 \\ a^2 & 0 & 0 & 0 & x & 0 & 0 & 0 & 1 \end{pmatrix} \qquad (8)$$

Using TMM, the One Site Density Matrix (OSDM) in the thermodynamic limit (TDL) can be obtained from the formula

$$\rho_1 = \frac{1}{\lambda_1}\langle e_1|f|e_1\rangle \qquad (9)$$

Where $\lambda_1$ is the largest eigenvalue of $F$ and $|e_1\rangle$ is the corresponding normalized eigenvector.

For $x = -3$, $\gamma = -2$, we obtain $\rho_1$ as a diagonal matrix in the basis $(|\bar{2}\rangle, |\bar{1}\rangle, |0\rangle, |1\rangle, |2\rangle)$ with the following diagonal elements

$$\rho_{11} = \frac{2a^2}{(5 + a^2 + \sqrt{81 - 6a^2 + a^4})(2 + \frac{1}{36}(3 - a^2 + \sqrt{81 - 6a^2 + a^4})^2)}$$

$$\rho_{22} = \frac{2a^2}{(5 + a^2 + \sqrt{81 - 6a^2 + a^4})(2 + \frac{1}{36}(3 - a^2 + \sqrt{81 - 6a^2 + a^4})^2)}$$

$$\rho_{33} = \frac{2(3 - a^2 + \sqrt{81 - 6a^2 + a^4})}{(5 + a^2 + \sqrt{81 - 6a^2 + a^4})(2 + \frac{1}{36}(3 - a^2 + \sqrt{81 - 6a^2 + a^4})^2)}$$

$$\rho_{44} = \frac{2(3 - a^2 + \sqrt{81 - 6a^2 + a^4})}{(5 + a^2 + \sqrt{81 - 6a^2 + a^4})(2 + \frac{1}{36}(3 - a^2 + \sqrt{81 - 6a^2 + a^4})^2)}$$

$$\rho_{55} = \frac{2(2 + \frac{1}{9}(3 - a^2 + \sqrt{81 - 6a^2 + a^4})^2)}{(5 + a^2 + \sqrt{81 - 6a^2 + a^4})(2 + \frac{1}{36}(3 - a^2 + \sqrt{81 - 6a^2 + a^4})^2)}$$

(10)

However, for $x = 0$, $a = 0$, $\gamma = 1$, $\rho_1$ has a much simpler form in the same basis, which is given by

$$\rho_1 = \begin{bmatrix} \frac{a^2}{2(1+a^2)} & 0 & 0 & 0 & 0 \\ 0 & 0 & 0 & 0 & 0 \\ 0 & 0 & 1 & 0 & 0 \\ 0 & 0 & 0 & 0 & 0 \\ 0 & 0 & 0 & 0 & \frac{a^2}{2(1+a^2)} \end{bmatrix}$$

(11)

GS fidelity is determined by calculating the magnitude of the overlay between the normalized GS wave functions $|\psi_0(\lambda)\rangle$ and $|\psi_0(\lambda+d\lambda)\rangle$ for closely spaced values of the Hamiltonian parameters $\lambda$ and $\lambda + d\lambda$ [28,42–44]. The following equation gives the definition of Global fidelity,

$$F(\lambda, \lambda + d\lambda) = |\langle\psi_0(\lambda)|\psi_0(\lambda+d\lambda)\rangle| \quad (12)$$

RFS is characterized as the intersection between the density matrices (of the subsystem) $\rho \equiv \rho(\lambda)$ and $\tilde{\rho} \equiv \rho(\lambda + \delta\lambda)$ of the GSs $|\varphi_0(\lambda)\rangle$ and $|\varphi_0(\lambda+\delta\lambda)\rangle$, where $\lambda$ and $\lambda + \delta\lambda$ represent two closely positioned values of the parameter $\lambda$. The RF is represented as

$$F_R(\lambda, \lambda+\delta\lambda) = \text{Tr}\sqrt{\rho^{1/2}\tilde{\rho}\rho^{1/2}} \quad (13)$$

We have studied the behaviour of OSVNE $S(i)$, its derivatives and the RF measures in the Spin-*2* chain system, especially around the Isotropic point (e.g., $x = -3$, $a = \sqrt{6}$, $\gamma = -2$) and degeneracy point ($x = -3$, $a = 0$, $\gamma = -2$), for the 'acritical' choice of the parameter values, and parallelly around the quantum critical point ($x = 0$, $a = 0$, $\gamma = 1$) of the said system.

## Results & Discussions

First, we have calculated and studied the variation of SSE and its derivatives with changing *a* for $x = -3$, $\gamma = -2$. The measure exhibits a maximum (the maximum value being $log_2 5$, for the VBS state of a spin-2 system) at the AKLT point $a=\sqrt{6}$ (*see figure (1)*) indicating enhancement of Quantum correlations captured by SSE (i.e., entanglement of one single spin-2 site with the rest of the system) up to the isotropic point, where the system crosses over to easy-plane anisotropy from easy-axis one. On the other hand, it attains a local minimum at the point $a = 0$, where the GS stops being unique and becomes finitely degenerate (degeneracy $\sim 3^L$). Interestingly, the double derivative of the same measure exhibits a local dip at the AKLT point and a singularity at the degenerate point $a=0$ (*inset of figure (1)*), even though the spin-2 chain system is 'non-critical' throughout the said parameter space. Both the entanglement measure and the correlation lengths are finite everywhere which establishes the fact that the Quantum correlation, captured by the measure under consideration, i.e., the entanglement of one single site with the rest of the system, is spread over finite length scales by finite amounts showing no critical behaviour whatsoever. However, the measure first increases with increasing up to the AKLT point, depicting enhancement of quantum correlations with decreasing easy-plane anisotropy, exhibits a maximum at the isotropic AKLT point and then again decreasing with the increasing easy-axis anisotropy (for values of a>√6).

We studied the variation of the same measure with varying *a* for $x = 0$, $\gamma = 1$. This particular choice of parameters makes the system critical at the point $a=0$, where the excitation energy gap vanishes, both the correlation lengths diverge, and the string order parameter, which has a non-zero value for the spin liquid state $a > 0$, vanishes. The OSVNE measure vanishes at the QCP $a=0$. It then increases very sharply with increasing *a* at first, and after exhibiting a maximum at the point a=√2 (which, by the way, does not correspond the to an isotropic AKLT state, and the maximum value of OSVNE is $log_2 3$, not $log_2 5$), at which the correlation lengths happen to be equal to each other and cross each other. It then decreases in a very slow manner with increasing *a*. Vanishing of the measure of quantum correlations at the QCP is a novel phenomenon unlike the conventional QCPs which usually tend to enhance and maximize correlations. In this case, apparently, the correlations are spread over larger length scales at the cost of its magnitude. The double derivative of the measure diverges at the transition point, which is a general feature of a continuous phase transition. The other peaks and dips exhibited by the 2[nd] derivative of the measure, however, demands more physical understanding.

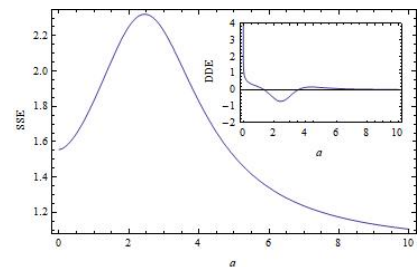

**Fig.1.** Plot of the single-site entanglement *S(i)* (SSE) and its double derivative, DDE (inset) with *a* for $x = -3$, $\gamma = -2$.

Next, we calculate the RF of a subsystem consisting of one single spin site $\rho(i)$. Inset of Figure (3) shows the variation of RF with the parameter *a* for *x* = -3, *γ* = -2. The measure shows sharp dip at the AKLT point indicating an abrupt change in the structure of the GS- although in this case, only a change in the nature of the anisotropy, not a quantum phase transition. The reduced Fidelity susceptibility (RFS) $\frac{\partial^2 F}{\partial a^2}$ shows a sharp peak at the AKLT point. This kind of behavior has usually been observed when the system undergoes a critical transition. Both RF and RFS is found to be insensitive to the degenerate point *a=0*.

For for *x* = 0, *γ* = 1, the behaviours of RF and RFS are different (see *figure 4*). The RF shows a sharp deep and the RFS shows a sharp peak at the QCP, *a=0*, successfully signalling a QPT. These measures, however, are completely insensitive to the point a=√2.

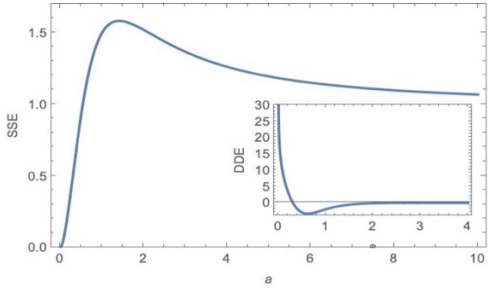

**Fig.2.** Variation of the single-site entanglement S*(i)* (SSE) and its double derivative, DDE (inset) with *a* for *x* = 0 , *γ* = 1.

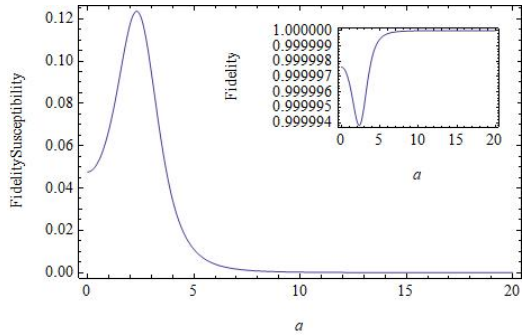

**Fig.3.** Plot of the Fidelity (inset) and Fidelity Susceptibility with *a* for *x* = -3, *γ* = -2.

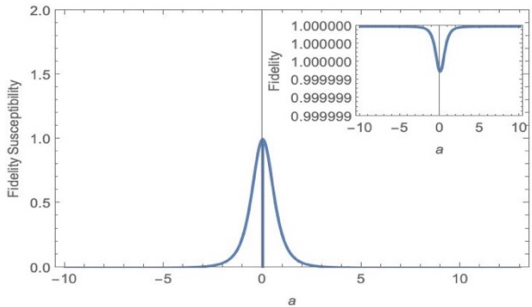

**Fig.4.** Variation of the Fidelity (inset) and Fidelity Susceptibility with *a* for *x* = 0 , *γ* = 1.

## Summary & Outlook

The entanglement and fidelity measures studied previously in a number of both existing and novel quantum systems have been proven to be efficient indicators of quantum critical transitions, across which those measure have shown non-trivial behaviours like exhibiting cusps, maxima, minima, discontinuity, and singularities with even specific scaling relations. In this work, we have found the same measures to behave nontrivially around points in the tuning parameter space which are not critical points, but either a cross-over point (i.e., a= $\sqrt{3}$, where the system is isotropic, and it goes over to easy-axis type anisotropy from easy-plane type as we increase the value of *a*) or a finitely 'degenerate' point (i.e., *a*=0, where the system becomes degenerate which grows ~ $3^L$ with the system size *L*).

On the other hand, for another choice of the parameters, when the system undergoes a continuous quantum phase transition from a string-ordered spin liquid phase to a 'not-string-ordered' disordered phase, the same entanglement measure becomes zero at the said point where the correlation lengths are infinite and the double derivative of the measure blows up, signaling a QPT successfully. It also exhibits a maxima at a point which is neither a QCP, nor an isotropic point. The RF and RFS measure, however, signals only the QCP in the conventional manner, and is insensitive to the the point where the OSVNE measure exhibits a maximum. The results reported here are partly analyzable physically and some results demands more physical understanding. The results, however rules out a single unified rule of consistency of the behaviours of the QIT measures in different physical systems and shed lights on a number of unexplored physical phenomena.

MPGSs have unique attributes that may be associated with their entanglement properties, symmetries, or other distinctive qualities [45,46]. There is a host of works which implies that there are other aspects to investigate concerning these captivating MPGSs [47]. Researchers may further explore the mathematical features, physical ramifications and practical uses of these states. These MPGSs encode quantum correlations and entanglement properties, contributing to our understanding of GS configurations in quantum systems. While the work in this paper highlights the most notable features, further details await publication, promising exciting exploration of these intriguing MPGSs' mathematical properties, physical implications, and applications [21].

The dynamics of trapped ions in a periodic potential, mapped to an AFM spin chain, showcases the versatility of experimental platforms for exploring frustration and interactions in spin-2 chains. Collectively, these studies offer a comprehensive understanding of the exact GS properties of spin-2 chains, leveraging various theoretical and experimental approaches to address the complexities of these systems. The result obtained in this work can be a typical or universal feature of similar quantum spin systems with different type of interactions, geometry, and phase diagrams. This work can be extended to a host of such physical systems, both, critical and non-critical, to obtain a general theoretical framework regarding behavior of entanglement measures and other QIT measures in physical systems. Novel measures, both operational and abstract, which can capture more complex structures of quantum correlations, e.g., multipartite entanglement can be designed and studied in these Quantum systems.

**References**

[1] Haldane FDM. 1983 Feb 14 Phys Lett A. ;93(9):464–8.

[2] Haldane FDM.1983 Apr 11 Phys Rev Lett. ;50(15):1153–6.

[3] Bethe H. 1931 Mar 1 Z Für Phys. ;71(3):205–26.

[4] White SR, Noack RM. 1992 Jun 15 Phys Rev Lett. ;68(24):3487–90.

[5] White SR. 1992 Nov 9 Phys Rev Lett. ;69(19):2863–6.

[6] Niggemann H, Zittartz J. 1996 Mar 1 Z Für Phys B Condens Matter. ;101(2):289–97.

[7] Klümper A, Schadschneider A, Zittartz J. 1993 Nov 1 Europhys Lett EPL. ;24(4):293–7.

[8] de Boer J, Schadschneider A. 1995 Dec 4 Phys Rev Lett. ;75(23):4298–301.

[9] Asoudeh M. 2012 Jan 1 Int J Theor Phys. ;51(1):246–58.

[10] Niggemann H, Klümper A, Zittartz J. 1997 Jan Z Für Phys B Condens Matter. ;104(1):103–10.

[11] Derrida B, Evans MR, Hakim V, Pasquier V 1993 Apr J Phys Math Gen. ;26(7):1493.

[12] Alipour S, Karimipour V, Memarzadeh L. 2008 Mar 1 Eur Phys J B. ;62(2):159–69.

[13] Alipour S, Baghbanzadeh S, Karimipour V. 2008 Dec EPL Europhys Lett. ;84(6):67006.

[14] Nakamura M, Nishimoto S. 2018 Sep 17 Eur Phys J B. ;91(9):203.

[15] Fannes M, Nachtergaele B, Werner RF. 1989 Dec Europhys Lett. ;10(7):633.

[16] Batchelor MT, Yung CM. 1994 Nov Int J Mod Phys B. ;08(25n26):3645–54.

[17] Tu HH, Sanz M. 2010 Sep 3 Phys Rev B. ;82(10):104404.

[18] Takano K, Suzuki H, Hida K. 2009 Sep 11 Phys Rev B. ;80(10):104410.

[19] Osterloh A, Amico L, Falci G, Fazio R. 2002 Apr Nature. ;416(6881):608–10.

[20] Osborne TJ, Nielsen MA. 2002 Sep 23 Phys Rev A. ;66(3):032110.

[21] Kim P, Katsura H, Trivedi N, Han JH. 2016 Nov 15 Physical Review B. 2016 Nov 15;94(19):195110.

[22] Wu LA, Sarandy MS, Lidar DA. 2004 Dec 15 Phys Rev Lett. ;93(25):250404.

[23] Costantini G, Facchi P, Florio G, Pascazio S. 2007 Jun J Phys Math Theor. ;40(28):8009.

[24] de Oliveira TR, Rigolin G, de Oliveira MC, Miranda E. 2006 Oct 23 Phys Rev Lett. ;97(17):170401.

[25] Tribedi A, Bose I. 2007 Apr 3 Phys Rev A. ;75(4):042304.

[26] Zhou HQ. 2007 Apr 23 arXiv preprint arXiv:0704.2945.

[27] Liu, J., Shi, Q., Zhao, J., & Zhou, H. (2009). *Journal of Physics A: Mathematical and Theoretical, 44*.

[28] Paunković N, Sacramento PD, Nogueira P, Vieira VR, Dugaev VK. 2008 May 1 Phys Rev A. ;77(5):052302.

[29] Gu SJ, Kwok HM, Ning WQ, Lin HQ. 2008 Jun 9 Phys Rev B. ;77(24):245109.

[30] Ma, J., Xu, L., & Wang, X. (2008). *arXiv: Quantum Physics*.

[31] Gu SJ. 2010 Sep 20 Int J Mod Phys B. ;24(23):4371–458.

[32] Tribedi A, Bose I. 2009 Jan 30 Phys Rev A. ;79(1):012331.

[33] Chandra AK, Das A, Chakrabarti BK. 2010 Springer Science & Business Media;313 p.

[34] Granroth GE, Meisel MW, Chaparala M, Jolicoeur T, Ward BH, Talham DR. 1996 Aug 19 Phys Rev Lett. ;77(8):1616–9.

[35] Ahrens MA, Schadschneider A, Zittartz J. 2002 Sep Europhys Lett EPL. ;59(6):889–95.

[36] Rigolin G, de Oliveira TR, de Oliveira MC. 2006 Aug 18 Phys Rev A. ;74(2):022314.

[37] Vidal G, Latorre JI, Rico E, Kitaev A. 2003 Jun 2 Phys Rev Lett. ;90(22):227902.

[38] O'Connor KM, Wootters WK. 2001 Apr 13 Phys Rev A. ;63(5):052302.

[39] Wootters WK. 1998 Mar 9 Phys Rev Lett. ;80(10):2245–8.

[40] Arnesen M, Bose S, Vedral V. 2001 June 14 Phys Rev Lett.;87, 017901

[41] Gunlycke D, Kendon VM, Vedral V, Bose S. 2001 Sep 6 Phys Rev A. ;64(4):042302.

[42] Cozzini M, Ionicioiu R, Zanardi P. 2007 Sep 19 Phys Rev B. ;76(10):104420.

[43] Zanardi P, Paunković N. 2006 Sep 26 Phys Rev E. ;74(3):031123.

[44] Quan HT, Song Z, Liu XF, Zanardi P, Sun CP. 2006 Apr 14 Phys Rev Lett. ;96(14):140604.

[45] Critch A, Morton J. 2014 Sep 10 Symmetry Integrability Geom Methods Applications SIGMA 10, 095

[46] Perez-Garcia D, Verstraete F, Wolf MM, Cirac JI. 2006 Aug 25 Quantum Inf. Comput., 7,401-430

[47] Davoudi Z, Mueller N, Powers C. 2023 Aug 21 Phys Rev Lett. ;131(8):081901.